# Metalized polymer tubes for high frequency electromagnetic waveguiding

*Dmitry Filonov[1,2,‡,=], Hahi Barhom[1,=], Andrey Shmidt[1], Yelena Sverdlov[1],*
*Yosi Shacham-Diamand[1], Amir Boag[1], Pavel Ginzburg[1,2]*

[1]*School of Electrical Engineering, Tel Aviv University, Tel Aviv, 69978, Israel*
[2]*Light-Matter Interaction Centre, Tel Aviv University, Tel Aviv, 69978, Israel*

[‡]*dimfilonov@gmail.com*
=contributed equally

**Abstract:** Low loss electromagnetic energy transport over long distances motivates the development of different types of waveguiding systems. Requirements of high quality optically polished waveguide surfaces needed in high-frequency applications and low-cost manufacturing are practically incompatible in current realizations. Here we demonstrate a new paradigm solution, based on surface functionalization with subsequent electroless plating of conductive micron smooth copper layer on the inner surface of flexible non-conducting poly-carbonate tubes. The structure was shown to support moderately low loss guiding performances (~5-10 dB/m) at Ku-band. The mechanically flexible design of the system allows shaping the waveguide network almost on demand. In particular, an efficient energy guiding over a closed loop with 8 lambda radius was demonstrated. The new platform of high quality metalized flexible waveguiding systems opens new opportunities in designs of cheap and efficient networks, operating over a broad spectral range, approaching tens of GHz and even higher.

*Introduction*

Efficient transport of electromagnetic energy over distances requires the development of different types of waveguiding systems. Many architectures have been demonstrated over the years and each one provides solutions for a specific frequency range and related applications[1]. Minimization of propagation and bending losses along with inerrability within larger-scale devices are among the key parameters, required from efficient interconnecting systems. Losses become a critical factor, affecting the performance of millimeter wave devices. Interface roughness on the level of optically polished high-quality surfaces is required from the implementations. As a result, the overall prices and bulky realizations of waveguiding components become a significant factor. Minimization of bending losses can be achieved by exploiting geometries, where electromagnetic radiation is enclosed within a confined volume. Typical representative examples here include rectangular and circular/elliptical geometries. Furthermore, the guided mode in those structures is confined in a void and, as a result, propagation losses are minimized (especially, if the core is vacuumed). Though some commercial solutions are available, bending of enclosed waveguides, is quite a complex technological challenge, as the millimeter-size aperture (Ka-band and higher) should be maintained along the entire curved trajectory.

Metallization of plastic components is fast developing technological direction, as it can provide significant advantages over conventional solutions. Several successful demonstrations of this approach include 3D printed metalized waveguide filters [2], waveguiding systems and related components [3], [4] (also for high GHz-THz and without metallization [5],[6]), antenna devices and components [7],[8],[9], and others. Furthermore, it is worth mentioning other additive manufacturing techniques, which allow the production of high-quality RF components, with an emphasis on antenna devices [10]. Different types of antennas, fabricated with CNC milling technique [11],[12], Laser Direct Structuring [13],[14],[15], conformal printing of metallic inks [16], conductive inkjet printing [17], ultrasonic wire mesh embedding [18], and metal deposition trough a mask on a curve surface [19], [20], were reported. Furthermore, integration of 3D printed plastic materials within antenna designs was demonstrated (e.g. [21],[22]) and fabrication of low-profile devices with several materials has been shown [23].

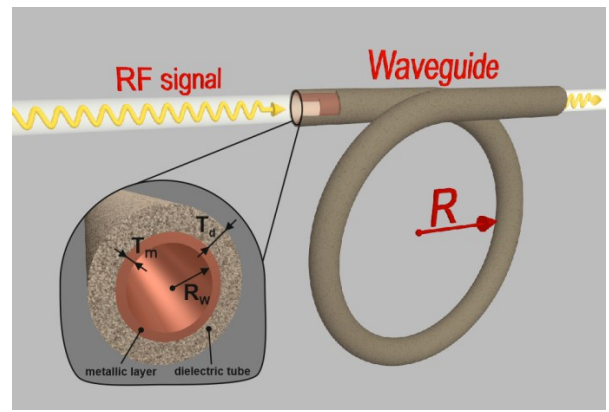

Fig. 1 Schematics of a metalized bendable dielectric tube for high frequency waveguiding applications.

In general, three main advantages of 3D printed polymers-based structures can be identified. The first one is their lightweights in comparison with solid metal made counterparts. Since only a thin conductive layer (skin depth of several microns for high GHz frequencies, if conventional metals are in use) is required for providing efficient waveguiding properties, lightweight polymers can serve as bulk rigid materials, supporting the structure. The second advantage of metalized 3D-printing approach to waveguiding systems is their potential flexibility to provide quite complex geometries (e.g. complex 3D networks, power divides, and others), which are hard to obtain with conventional milling techniques. It is worth noting that rigid 3D structures, once being fabricated, cannot be reshaped. Our proposal, however, is lacking of these limitations. The last and already mentioned factor is the manufacturing cost, which can be significantly reduced in the case when additive manufacturing techniques are in use.

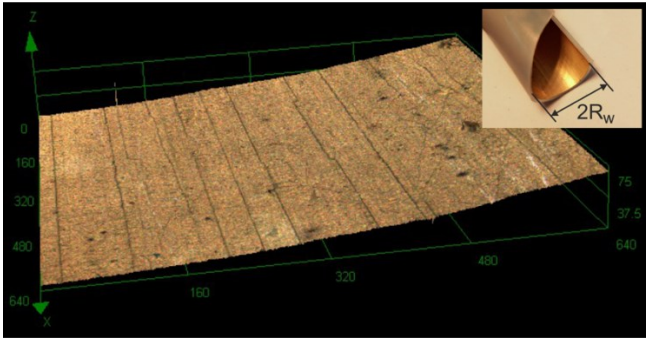

Fig. 2 Surface profile of 2.5 mm × 1 mm section of a metalized tube. (Inset) Photograph of the cut section of the tube.

Here we demonstrate a new waveguiding system, based on metallization of flexible polymer tubes (Fig. 1). The distinctive advantages of this architecture include its extremely low cost and mechanical flexibility. Inner surfaces of initially bendable polymer tubes are metalized with an electroless plating technique and, as a result, high-quality RF conductivity is achieved. Furthermore, controllable chemical deposition process was demonstrated to provide high-quality metallization layers with micron-scale roughness along centimeter-range tube cross-sections. Mechanical flexibility of the tube waveguides allows bending them almost on demand without requiring an a priori knowledge of a layout of an end-user.

The Letter is organized as follows: basic technological principles of the metallization are discussed first and then followed by experimental demonstration of the new system and its performance, which is compared with standard existing solutions.

*Tubes metallization*
Tubes with different polymeric composites can be metalized by electroless deposition of a copper layer. As a first step, surface of the tubes was washed from excess or leftover materials that stay after chemical treatment and pre-production post-processing. The tube was connected to a homemade peristaltic pump, which maintains a high throughput flow of functional solutions through it. The cleaning cycle is applied at the end of the metallization process. DIW (Di Ionized Water) wash at room temperature was followed with methanol absolute dry wash at 50°C for 5 minutes followed by rinsing with DIW. Then N,N-Dimethylformamide solution at 50°C washed the tube for 5 minutes followed by rinsing with Ethanol then DIW. Etching to reduce the roughness on the tube walls by Chromo-Phosphoric-Sulfuric acid at 50°C for 5 minutes was performed and followed by DIW rinsing. The next step is a sensibilization in a solution, containing SnCl2-70 g/l and HCl – 40 ml/l for 30 min at room temperature, followed by DIW rinsing. Pd-activation for 60 min at room temperature with PdCl2 1g/l solution, followed by DIW rinsing. Then copper solution 15 g/l, K-Na-tartrate 30 g/l $Na_2CO_3$ 10 g/l NaOH 40 g/l Formaldehyde 35%. Adjusting pH in the range 12.5 – 12.7 assuring metal layer formation properly at a proper rate compared to the reduction time of copper.

The surface roughness of copper layer was measured in Olympus LEXT OLS4100 laser scanning, providing a resolution of 10-20 nm. The quality of the surface was defined as the standard deviation of points on the surface from the mean position. The measurement was performed over $mm^2$-scale area. The surface roughness was estimated to be around 10μm, while smaller areas of investigation were found to be as smooth as ± 0.2μm. It indicates that more accurate and repetitive surface cleaning can allow obtaining those numbers (corresponding to surface qualities of current commercially available 60 GHz hollow waveguide systems). The overall thickness of the metal layer, deposited on the inner side of the tube, is few microns (depending on copper solutions flow and duration), while the electromagnetic skin depth in copper at 10 GHz frequency is less than a micron. Fig. 2 (inset) shows a photo of a cut section of the metalized tube. The topographic image of the surface appears in the main figure (Fig. 2). The slant of the surface corresponds to the physical profile behavior, which can be almost perfectly fitted with a quadratic 2D polynomial. The quasi-periodic strips of the height profile (if projected on the flat landscape) are observed along the direction of the stream, which was induced inside the tube during the metallization process.

*Straight waveguide sections*
A metalized tube section was assessed for the waveguiding at the next stage and compared with a straight brass duct section. Standard SMA to waveguide connectors were implemented on short sections of brass ducts (inner diameter of 13 and 11mm in the case of brass and metalized sections, respectively), sealed from the outer side (insets to Fig. 3).

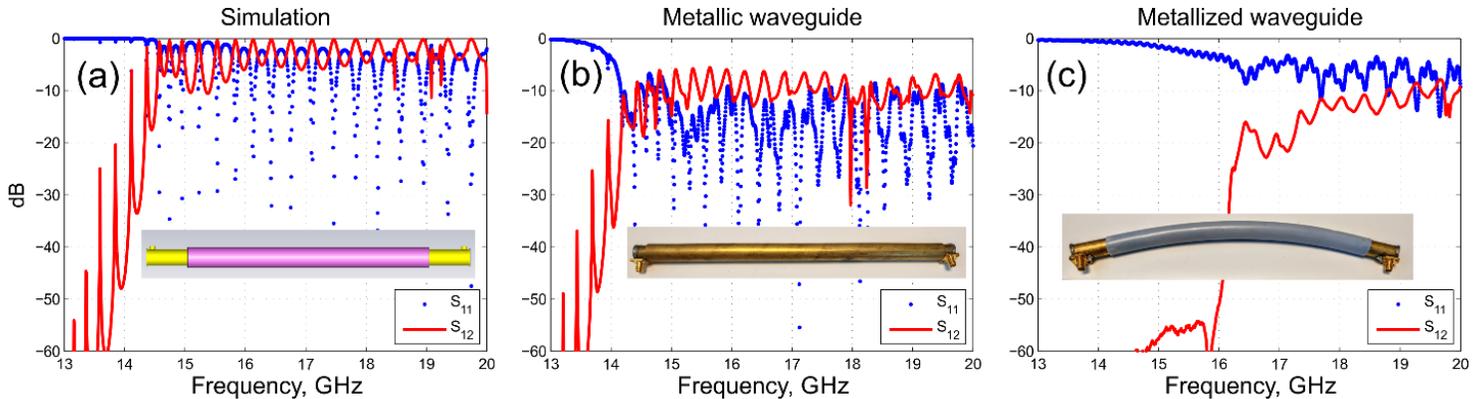

Fig. 3 S-parameters (absolute values) of circular waveguide system ($S_{11}$ – blue dots, $S_{12}$ – red solid lines). (a) Numerical simulation. (b) Straight section of a brass duct (inner diameter is 13mm). (c) Metalized polymer tube (inner diameter is 11 mm). Sections are 26cm in length.

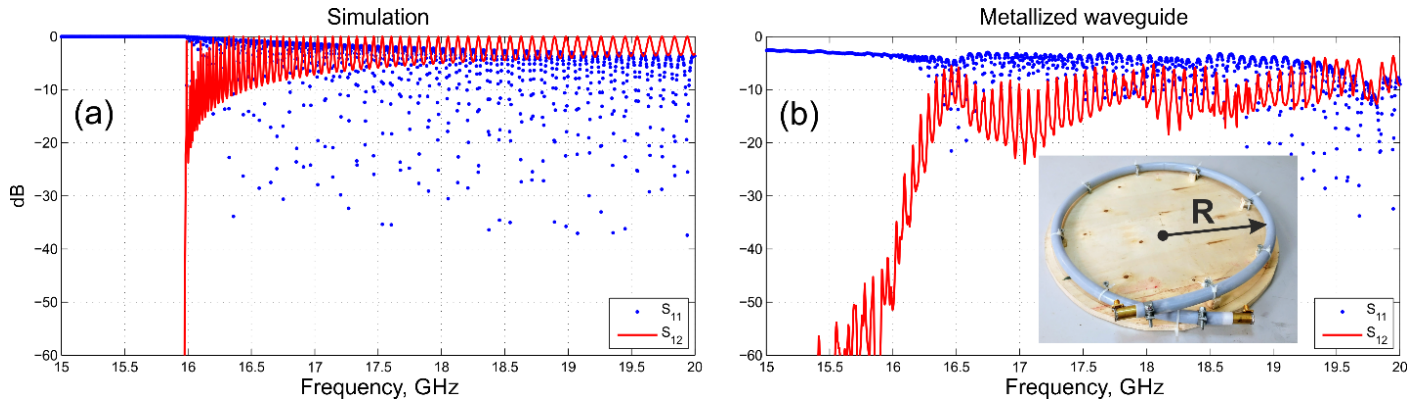

Fig. 4 S-parameters (absolute values) of the bended waveguide system (S11 – blue dots, S12 – red solid lines). (a) Numerical simulation. (b) Metalized polymer tube (11 mm inner diameter, 1 meter length, 16 cm bending radius).

Transmission and reflection coefficients (S-parameters) of the system were retrieved with the VNA (N5232B PNA-L Microwave Network Analyzer) after performing calibration procedures (Fig. 3). Numerical simulation of perfectly conducting (PEC) tube section was performed with CST microwave studio (frequency domain solver). Custom-made connectors were explicitly taken into account, while the excitation/collection ports were placed directly on the coaxial cables. As a result, imperfect mode matching was emulated, replicating the experimental scenario. The cut-off frequency of the waveguide with circular cross-section has a well known analytical expression. Substitution of the tube's parameters leads to the cut-off at ~13.5 GHz, which agrees well with the numerical results.

Fig. 3(b) is the reference measurement, demonstrating the performances of an etalon structure – a brass duct. The results agree well with the numerical data with certain deviations, including the absolute values of S-parameters, which correspond to the leakage of the radiation via connection ports. This effect is more pronounced for higher frequencies, where electromagnetic fields squeeze through gaps between the connectors, which are imperfectly attached to the duct's surface. The performance of the metalized tube appears on Fig. 3(c). The structure has got a slight distortion during the metallization process. This can be prevented in the future by mechanical stabilization of the whole section during either pre- or post-processing. The direct comparison between the brass (etalon) and metalized sections show remarkable similarities, nevertheless, the overall amount of loss is 2-6 dB higher. Also, the cut-off frequency is shifted to ~16GHz, since the inner diameter of the metallized polymer section is 11mm, which corresponds to this number.

*Bended waveguides*

The distinctive advantage of the flexible waveguide platform is to support bending geometries - Fig. 4 (b, inset). Here, the overall length of the metalized 11mm diameter tube is 1m, while the bending radius is 16 cm. Fig. 4(a) shows the result of the numerical simulation, carried out under the same conditions, as for the straight section (the PEC duct + connectors). The cut-off frequency is clearly observed at 16 GHz and it is more pronounced in this case since this bent waveguide section is 3 times longer than the straight section (Fig. 3). Fig. 4(b) shows the experimental results, which correspond well to the numerical estimate. Similarly to the case of the straight section, the losses of the real structure are higher by ~6 dB, which is consistent with the qualitative comparison, which was performed on the data from Fig. 3.

*Conclusions*

A new platform for cheap and flexible high-frequency guiding was proposed and experimentally demonstrated. The principle is based on electroless plating of smooth bendable polymer tubes, whose inner surfaces undergo metallization. Highly conductive chemically deposited layers demonstrate properties, comparable with optically polished surfaces of conventional brass-based waveguide sections. As a result, efficient waveguiding at Ku-band was demonstrated. It is worth noting, that the developed methodology can be extended to higher frequencies, approaching 100 GHz and even higher. Standard waveguiding solutions demonstrate excessive losses at those frequencies. Furthermore, waveguide components become extremely expensive, preventing high-frequency RF technologies to replace existing widespread low GHz solutions (e.g. Wi-Fi and many others). Our solution might suggest a paradigm shift in this area.

Acknowledgments
The research was supported in part by PAZY Foundation, 3PEMS Ltd.

References:

[1]  D. M. Pozar, *Microwave engineering*. Wiley, 2012.

[2]  E. Massoni, M. Guareschi, M. Bozzi, L. Perregrini, U. A. Tamburini, G. Alaimo, S. Marconi, F. Auricchio, and C. Tomassoni, "3D printing and metalization methodology for high dielectric resonator waveguide microwave filters," in *2017 IEEE MTT-S International Microwave Workshop Series on Advanced Materials and Processes for RF and THz Applications (IMWS-AMP)*, 2017, pp. 1–3.

[3]  M. I. M. Ghazali, K. Y. Park, V. Gjokaj, A. Kaur, and P. Chahal, "3D Printed Metalized Plastic Waveguides for Microwave Components," *Int. Symp. Microelectron.*, vol. 2017, no. 1, pp. 000078–000082, Oct. 2017.

[4]  E. A. Rojas-Nastrucci, J. T. Nussbaum, N. B. Crane, and T. M. Weller, "Ka-Band Characterization of Binder Jetting for 3-D Printing of Metallic Rectangular Waveguide Circuits and Antennas," *IEEE Trans. Microw. Theory Tech.*, 2017.

[5]  M. Weidenbach, D. Jahn, A. Rehn, S. F. Busch, F. Beltrán-

Mejía, J. C. Balzer, and M. Koch, "3D printed dielectric rectangular waveguides, splitters and couplers for 120 GHz," *Opt. Express*, vol. 24, no. 25, p. 28968, Dec. 2016.

[6] D. W. Vogt, J. Anthony, and R. Leonhardt, "Metallic and 3D-printed dielectric helical terahertz waveguides," *Opt. Express*, vol. 23, no. 26, p. 33359, Dec. 2015.

[7] B. Zhang, P. Linner, C. Karnfelt, P. L. Tarn, U. Sodervall, and H. Zirath, "Attempt of the metallic 3D printing technology for millimeter-wave antenna implementations," in *2015 Asia-Pacific Microwave Conference (APMC)*, 2015, vol. 2, pp. 1–3.

[8] E. G. Geterud, P. Bergmark, and J. Yang, "Lightweight Waveguide and Antenna Components Using Plating on Plastics," *7th Eur. Conf. Antennas Propogation*, 2013.

[9] R. Zhu and D. Marks, "Rapid prototyping lightweight millimeter wave antenna and waveguide with copper plating," in *IRMMW-THz 2015 - 40th International Conference on Infrared, Millimeter, and Terahertz Waves*, 2015.

[10] M. Liang, J. Wu, and X. Yu, "3D printing technology for RF and THz antennas," *Ieee*, no. 2, pp. 536–537, Jan. 2016.

[11] M. Ferrando-Rocher, J. I. Herranz, A. Valero-Nogueira, and B. Bernardo, "Performance Assessment of Gap Waveguide Array Antennas: CNC Milling vs. 3D Printing," *IEEE Antennas Wirel. Propag. Lett.*, vol. 1225, no. c, pp. 1–1, 2018.

[12] M. A. Al-Tarifi and D. S. Filipovic, "On the design and fabrication of W-band stabilised-pattern dual-polarised horn antennas with DMLS and CNC," *IET Microwaves, Antennas Propag.*, vol. 11, no. 14, pp. 1930–1935, Nov. 2017.

[13] F. Sonnerat, R. Pilard, F. Gianesello, F. Le Pennec, C. Person, and D. Gloria, "Innovative LDS Antenna for 4G Applications," *IEEE*, no. Eucap, pp. 2696–2699, Apr. 2013.

[14] A. Friedrich, M. Fengler, B. Geck, and D. Manteuffel, "60 GHz 3D integrated waveguide fed antennas using laser direct structuring technology," in *2017 11th European Conference on Antennas and Propagation, EUCAP 2017*, 2017, pp. 2507–2510.

[15] A. Friedrich and B. Geck, "On the Design of a 3D LTE Antenna for Automotive Applications based on MID Technology," *Eur. Microw. Conf.*, pp. 640–643, 2013.

[16] J. J. Adams, S. C. Slimmer, T. F. Malkowski, E. B. Duoss, J. A. Lewis, and J. T. Bernhard, "Comparison of Spherical Antennas Fabricated via Conformal Printing: Helix, Meanderline, and Hybrid Designs," *IEEE Antennas Wirel. Propag. Lett.*, vol. 10, pp. 1425–1428, 2011.

[17] M. Ahmadloo, "Design and fabrication of geometrically complicated multiband microwave devices using a novel integrated 3D printing technique," in *2013 IEEE 22nd Conference on Electrical Performance of Electronic Packaging and Systems*, 2013, pp. 29–32.

[18] M. Liang, C. Shemelya, E. MacDonald, R. Wicker, and H. Xin, "3-D Printed Microwave Patch Antenna via Fused Deposition Method and Ultrasonic Wire Mesh Embedding Technique," *IEEE Antennas Wirel. Propag. Lett.*, vol. 14, pp. 1346–1349, 2015.

[19] I. Ehrenberg, S. Sarma, T. Steffeny, and B.-I. Wuy, "Fabrication of an X-Band conformal antenna array on an additively manufactured substrate," in *2015 IEEE International Symposium on Antennas and Propagation & USNC/URSI National Radio Science Meeting*, 2015, pp. 609–610.

[20] B.-I. Wu and I. Ehrenberg, "Ultra conformal patch antenna array on a doubly curved surface," in *2013 IEEE International Symposium on Phased Array Systems and Technology*, 2013, pp. 792–798.

[21] M. Mirzaee, S. Noghanian, and I. Chang, "Low-profile bowtie antenna with 3D printed substrate," *Microw. Opt. Technol. Lett.*, vol. 59, no. 3, pp. 706–710, Mar. 2017.

[22] C. Shemelya, M. Zemba, M. Liang, X. Yu, D. Espalin, R. Wicker, H. Xin, and E. MacDonald, "Multi-layer archimedean spiral antenna fabricated using polymer extrusion 3D printing," *Microw. Opt. Technol. Lett.*, vol. 58, no. 7, pp. 1662–1666, Jul. 2016.

[23] P. Parsons, M. Mirotznik, P. Pa, and Z. Larimore, "Multi-material additive manufacturing of embedded low-profile antennas," *Electron. Lett.*, vol. 51, no. 20, pp. 1561–1562, Oct. 2015.